# Highly coherent phase-lock of an 8.1 μm quantum cascade laser to a turn-key mid-IR frequency comb

B. Chomet,[1] D. Gacemi,[1] O. Lopez,[2] L. Del Balzo,[1] A. Vasanelli,[1] Y. Todorov,[1] B. Darquié,[2] and C. Sirtori,[1]

[1]*Laboratoire de Physique de l'Ecole normale supérieure, ENS, Université PSL, CNRS, Sorbonne Université, Université Paris Cité, Paris, France*

[2]*Laboratoire de Physique des Lasers, CNRS, Université Sorbonne Paris Nord, 93430 Villetaneuse, France*

A continuous-wave Fabry-Perot quantum cascade laser (QCL) emitting at 8.1 μm operating in single mode regime has been coherently phase locked to a turn-key low-noise commercial mid-Infrared (mid-IR) optical frequency comb. The stability of the comb used as a reference is transferred to the QCL resulting in an integrated residual phase error of 0.4 rad. The laser linewidth is narrowed by more than two orders of magnitude reaching 300 Hz at 1ms observation time, limited by the spectral purity of the mid-IR comb. Our experiment is an important step towards the development of both powerful and metrology-grade QCLs as well as fully-stabilized QCL frequency comb, and opens perspectives for precision measurements and frequency metrology in the mid-IR.

## 1. Introduction

High-spectral-purity narrow-linewidth continuous-wave (cw) lasers are crucial for both fundamental research and technological applications in many fields of science, such as atomic, molecular, quantum and optical physics. They play in particular a major role in precision measurement physics, in the development of quantum technologies, and in time and frequency metrology. For these applications, the free-running linewidth determined by the laser short-term stability is often not adequate and active frequency stabilization is required. In the near-infrared and visible domain, this is routinely achieved by locking the laser to the resonance of a high-finesse Fabry-Perot cavity, or to a narrow atomic or molecular resonance.

Frequency-stabilized sources in the mid-infrared (mid-IR, from 3 to 20 μm) domain are also needed, in particular for conducting ultra-high resolution molecular spectroscopy measurements in the so-called molecular fingerprint region with far-reaching applications ranging from fundamental physics [1–6] and metrology [7–9] to astrophysics, remote sensing and Earth sciences[10,11]. Quantum cascade lasers (QCLs) are a robust technology that cover the mid-IR spectrum, offer wide tuning capabilities and have cw output powers as high as several watts but QCLs show substantial frequency fluctuations [12–19] . Phase/frequency stabilized QCLs have been demonstrated [20], either by phase-locking to a $CO_2$ laser [21,22], frequency locking to a sub-Doppler molecular transition [23], optical injection locking [24] or phase-locking to an optical frequency comb, either calibrated to a commercial Rb frequency standard or H-Maser steered to a global navigation satellite system (GNSS) [18,19,25–29], or stabilized to a near-infrared metrology-grade frequency reference traceable to primary frequency standards [30–32] . References [31,32] report ultimate sub-Hz stabilities and accuracies, and linewidth narrowing down to the 0.1 Hz level. Locking to a frequency comb typically necessitates sophisticated non-linear optical setups requiring very delicate alignment procedures, and often penalizing the QCL output power, most of it being required for generating non-linear signals.

In this Letter, we measure the free-running frequency noise power spectral density of an 8.1 μm Fabry-Perot QCL operating in single mode regime and demonstrate coherent phase locking of this source to a tooth of a commercially available low-noise mid-IR optical frequency comb synthesizer (OFCS), by processing a beat-note signal generated directly in a mid-IR photodetector. This allows the spectral performances of the OFCS to be transferred to the QCL without the need to pass the QCL beam through a complex dedicated non-linear optical setup, thus allowing less than 1% of its optical power to be used for locking, keeping most of it at disposal for further experiments. Locking to such a low-noise comb results in a linewidth narrowing of the QCL by of factor of 250, from ~100 kHz down to the few hundreds of hertz level. This corresponds to a 2 orders of magnitude linewidth improvement compared to reference [25] that reports phase locking of a distributed feedback QCL to a home-built mid-IR comb. Locking the OFCS to a remote near-infrared ultra-stable frequency reference traceable to the atomic fountain clocks of the French metrology institute, as demonstrated in reference [31,32] will eventually provide a both powerful and metrology-grade QCL to be used in different applications such as high precision spectroscopy, quantum optics measurements or LIDAR (light detection and ranging).

## 2. Experiment

The experimental setup is shown in Fig. 1(a). A turn-key offset-free mid-IR OFCS from Menlo Systems, spanning the 7.4 to 9.4 μm spectral window (full-width-half-maximum FWHM ~4.29THz), is used as an optical reference. The mid-infrared comb is generated by difference frequency mixing in a non-linear crystal between a low noise 1.5 μm Erbium doped fiber frequency comb and its extension in a wavelength shifting fiber towards the 1.8 μm window. The mid-IR comb delivers an average output power of 2.2 mW, resulting in tens of nanoWatts comb modes. The comb repetition rate, $f_{\rm rep} = 100\ MHz$, is stabilized against a radio-frequency (RF) synthesizer using the electronic control unit of the comb. This corrects the slow frequency drift without affecting the laser linewidth. The QCL is a Fabry-Perot source grown and processed at ETH Zürich (more details about this laser are given in the supplementary material). It is typically operated close to threshold at a temperature of 263 K (threshold current is 400 mA), and delivers up to 45 mW in a continuous wave single mode oscillating at a wavelength of ~8.16 μm. Fig. 1(b) shows the optical spectrum of the mid-IR comb that overlaps with the single longitudinal mode of the QCL. The QCL and the mid-IR-OFCS beams are superimposed by means of a 70/30 germanium beam splitter. Both beam splitter outputs are optically filtered with a 50 GHz resolution monochromator, and focused onto an 800-MHz bandwidth mercury-cadmium-telluride (MCT) detector (VIGO System PVI-4TE-10.6) (see supplementary material). The QCL beam is attenuated and a total power of about only 100 µW is hitting the detector. The resulting beat-note signals obtained on detector 1 and 2 in Fig. 1(a) oscillate at:

$$f_{\text{beat}} = \nu_{\text{QCL}} - n \times f_{\text{rep}}$$

with $\nu_{QCL}$ the QCL frequency and n~ 370 000 the tooth number of the mid-IR-OFCS. Note that in our case the carrier-envelope offset $f_{CEO} = 0$, as the mid-IR comb is generated by frequency difference. Fig. 1(c) shows the RF spectrum of the signal measured at the output of detector 1 when the QCL operates in a free-running regime. Four main peaks emerge from the noise between DC and 150 MHz: the beat-note signal between the QCL mode and the nearest comb tooth at $f_{beat} \simeq 23$ MHz, a peak at the comb repetition rate of 100 MHz, and the beat-note signals between the QCL mode and the second- and third-nearest comb teeth at $f_{rep} \pm f_{beat}$ ($\simeq$77 MHz and $\simeq$ 123 MHz). The OFC/QCL beat signals are observed with a signal-to-noise ratio (carrier power to phase noise power density ratio) of SNR = 26 dB at 1MHz resolution bandwidth and are processed to either measure the frequency noise of the free-running QCL, or for phase-locking it to the OFCS. To characterize the residual phase fluctuations of the locked QCL, a second out-of-loop detector (detector 2) is used.

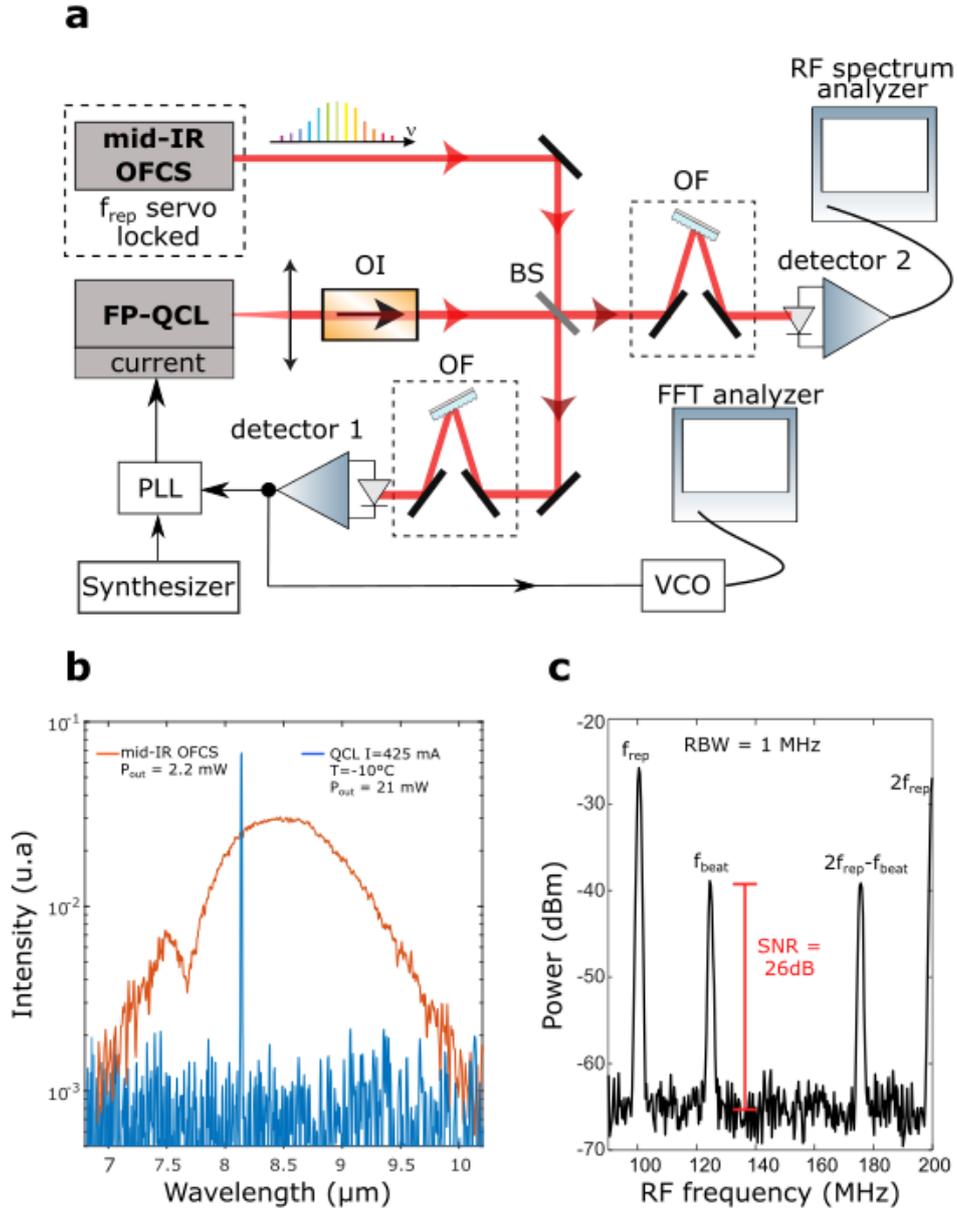

FIG. 1. (a) Experimental setup for measuring the frequency noise of the free-running QCL and for coherently phase-locking it to a low-noise mid-IR OFCS. FP-QCL, Fabry-Perot Quantum Cascade Laser; mid-IR OFCS, mid-infrared-optical frequency comb synthesizer; BS, beam splitter; OF, optical filtering monochromator build from a diffractive grating with 150 grooves/mm; PLL, phase lock loop; OI, optical isolator; VCO, voltage controlled oscillator; FFT, fast Fourier transform (b) Optical spectra of the FP-QCL and the mid-IR OFCS recorded with a 2 GHz resolution optical spectrum analyzer (the modes of the 100 MHz repetition rate OFCS are not resolved). (c) RF spectrum of the beat signal between the mid-IR OFCS and the free running QCL (1 MHz resolution bandwidth, RBW).

## 3. Free-running Fabry-Perot QCL frequency noise

The frequency noise power spectral density (FNPSD) of the free-running QCL is derived from the beat-note signal $f_{beat}+ f_{rep}$ measured on detector 1, in open-loop configuration. The latter is fine tuned to 140MHz with the QCL current ($\Delta\nu_{QCL}/\Delta I$ = -235MHz/mA) to match the frequency range of a tracking oscillator consisting of a voltage-controlled oscillator (VCO) phase-locked to the 140 MHz beat-note signal. The VCO tuning voltage then mirrors the beat-note signal's frequency noise and thus the free-running QCL frequency noise for Fourier frequencies below the system bandwidth of ~4 MHz. To keep the beat signal frequency within the tracking range, slow drifts of the free-running QCL (few MHz/s) produced by thermal and mechanical fluctuations have to be eliminated. This is achieved by implementing a slow feedback loop (sub-Hz bandwidth) on the QCL current to keep the beat-note at 140 MHz. The power spectrum of the VCO tuning voltage is then measured using a fast Fourier transform (FFT) analyzer, and the FNPSD of the QCL is finally derived using the 4 MHz/V VCO voltage-to-frequency conversion factor.

Fig. 2(a) shows the resulting FNPSD of the free-running QCL/mid-IR comb beat-note signal (blue line (I)). It has a 1/f trend at low frequency up to ~300 kHz. Above 300 kHz, the frequency noise increases up to approximately 4 MHz, beyond which we observe a decrease due to the finite tracking oscillator bandwidth. The increase of the frequency-noise above 300 kHz is the consequence of the finite beat-note signal-to-noise ratio of 86 dB/Hz, limited by our detector background noise (see supplementary material). The corresponding white phase noise around the optical carrier gives rise to a double-sided contribution of $(2/SNR_{1Hz}) \times f^2$ in our FNPSD measurement, with $SNR_{1Hz}$ the SNR in a 1 Hz bandwidth. To validate this, the black curve (IV) in Fig. 2(a) shows the FNPSD obtained by blocking the mid-IR-OFCS beam and by replacing $f_{beat}$ with a low phase noise artificial signal yielding the same $SNR_{1Hz}$ generated by modulating the QCL current with a RF generator. The flat plateau in the range 1 Hz-100 kHz is due to the FFT-analyzer background noise, and above 100-kHz the effect of the SNR can be seen as it matches the blue curve (I). The grey line (II) in Fig. 2(a) shows the contribution of the low-noise current source, obtained by multiplying the measured driver's current noise spectrum (~300 pA/Hz$^{1/2}$ above 100Hz) by the laser DC current-to-frequency response (-235 MHz/mA). Finally, the red dashed line (III) in Fig. 2(a) gives the contribution from the OFCS tooth frequency noise to the heterodyne signal. The FNPSD of this tooth at frequency $n \times f_{\text{rep}}$ is determined from typical noise measurements of the repetition rate of the OFCS 1.55 µm main oscillator provided by Menlo Systems. Note that this estimation does not consider the frequency noise that could be added by non-linear processes involved in the generation of the mid-IR OFCS which is assumed to be negligible.

For Fourier frequencies below 300 kHz, the blue curve (I) in Fig. 2(a) corresponds to the free-running QCL FNPSD, all other contributions being negligible. The level of frequency noise measured is similar to that reported in ref [17], another measurement carried out on a Fabry-Perot QCL. Yet, the expected white frequency noise level corresponding to the QCL Schawlow-Townes limit does not seem to be reached above 300 kHz due to our SNR value. However, an upper limit of $10^3$ Hz$^2$/Hz (and a corresponding upper limit of ~3 kHz on the intrinsic laser linewidth) can be inferred. The free-running QCL emission line shape is calculated from the measured FNPSD for a 1 ms integration time (IT) following [33] and shows a FWHM $\Delta\nu_{,QCL}$ ~72 kHz (see Fig. 2(a) inset). A similar calculation based on curve (III) in Fig. 2(a) leads to a linewidth $\Delta\nu_{,OFCS}$ = 300 Hz (1ms IT) for the OFCS tooth, once again confirming the negligible contribution of the OFCS noise in our measurement. Fig. 2(b) shows the evolution of $\Delta\nu_{,QCL}$ calculated for different integration times. It decreases with decreasing integration time and reaches a minimum of 15 kHz at 200 µs IT. For the shortest integration times (< 100 µs), the linewidth is Fourier limited: the contribution from the frequency noise is lower than the Fourier uncertainty.

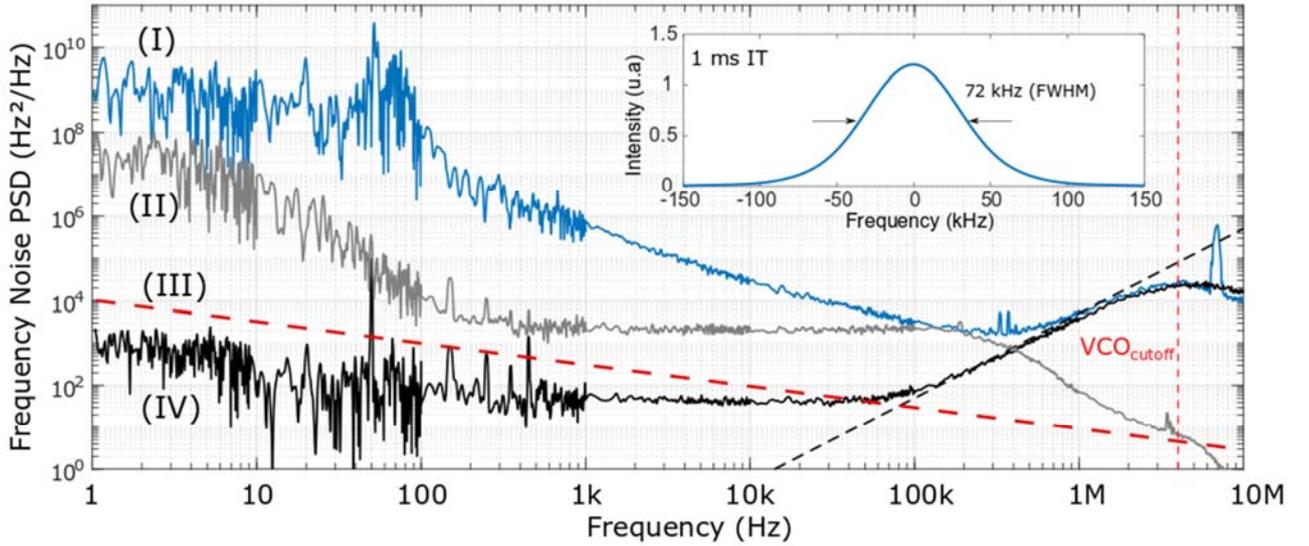

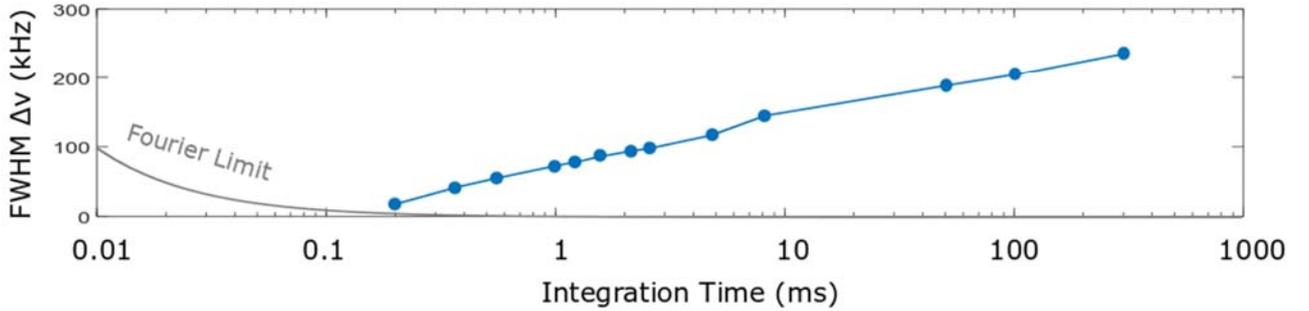

FIG. 2. (a) Frequency noise power spectral density (FNPSD) of the free-running 8.1 µm QCL/mid-IR comb beat-note signal (blue line (I)). The contributions from the laser driver current noise (grey line (II)) and from the mid-IR OFCS tooth contributing to the beat-note signal (red dashed line (III)) are also plotted for comparison. The black line (IV) corresponds to our noise floor, see text (with the black dashed line corresponding to the contribution from the limited beat-note signal-to-noise ratio of 86 dB/Hz). The tracking oscillator bandwidth is also shown (vertical red dashed line). The inset shows the free-running QCL emission line shape calculated from the measured FNPSD (I) for a 1 ms integration time (IT). (b) Blue line: free-running QCL FWHM emission line width as a function of integration time calculated from FNPSD (I) (the slight excess of noise of technical origin visible on curve (I) between 40 Hz and 100 Hz has been purposely removed for this calculation); the grey line corresponds to the Fourier limit resulting from the resolution bandwidth.

**4. Fabry-Perot QCL phase locking and linewidth narrowing**

As illustrated in Fig. 1(a), to coherently phase-lock the QCL to a tooth of the mid-IR-OFCS, a phase-error signal is generated by comparing the phase of the filtered and amplified beat signal with a reference signal typically at 140 MHz from a synthesizer (Zurich Instrument UHFLI). A phase-lock servo loop is used to apply a correction signal directly to the QCL's current. Note that much less than 1% of the QCL power (<100μW) is used for locking, keeping most of it at disposal for different experiments

In closed-loop operation, a significant narrowing of the beat-note signal linewidth is observed and a maximum signal-to-noise ratio of about 55 dB at 1 kHz resolution-bandwidth was obtained, as shown in the inset of Fig. 3. The beat signal spectrum is recorded on a RF spectrum analyzer after detection with a second out-of-loop photodetector (detector 2 in Fig. 1(a)), similar to the one in-loop, in order to avoid errors brought by the detection setup and associated electronics. A ~250 kHz phase-lock feedback bandwidth can be inferred from the spectral shape. Figure 3 shows the measured phase noise PSD of the beat signal between the Mid-IR-OFCS and the phase locked QCL. The latter is generated by comparing the phase of the beat signal at the output of detector 2 with a synthesized reference signal at 140 MHz (UHFLI Zurich Instrument) and processed with a FFT spectrum analyzer. A 0.39-rad rms residual phase noise is obtained in a 1 Hz to 1 MHz integration bandwidth, indicating that more than 85% of the power is concentrated in the coherent carrier of the beat signal [34].

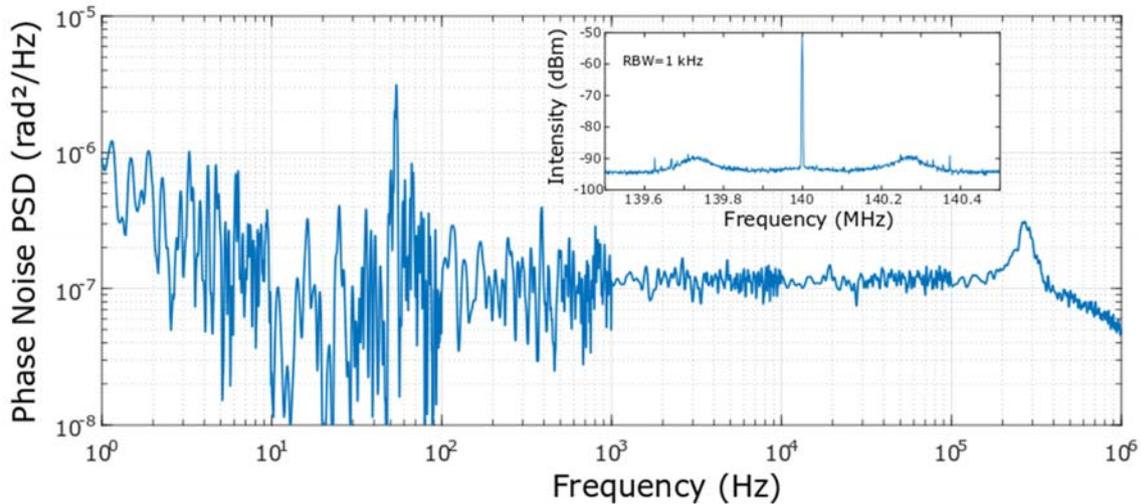

FIG. 3. Phase-noise power spectral density (PSD) of the beat signal between the phase-locked QCL and the mid-IR comb. Inset: Beat signal spectrum between the mid-IR OFC and the phase-locked QCL taken with a RF spectrum analyzer (1 kHz resolution bandwidth, RBW).

This is the signature of a highly coherent phase-lock and of the very good transfer of the mid-IR OFCS spectral features to the QCL. Therefore, this results in a QCL linewidth ultimately set by the mid-IR comb tooth linewidth. Exploiting the OFCS FNPSD displayed in Fig.2 (dashed red line (III)), we calculate a linewidth of respectively ~300 Hz (see above) and ~800 Hz at 1 ms and 100 ms integration time, which is 2 to 3 orders of magnitude lower than in the free-running regime. Our phase-locked Fabry-Perot QCL shows a 2 orders of magnitude narrower linewidth compared to a similar work using a 8.6 µm DFB QCL phase-locked to a home-built 250 MHz mid-IR comb [25].

## 5. Conclusion

We characterized the frequency noise of an 8.1 µm free-running Fabry-Perot QCL operating in single mode regime. It was then phase-locked to a low-noise commercial mid-infrared comb, itself generated from a near-infrared comb. We achieved a QCL linewidth reduction from ~100 kHz to the few 100 Hz level, limited by the linewidth of the mid-IR comb used as a reference, improving on previous studies. Locking the comb to an ultra-stable frequency reference calibrated to the primary frequency standards of the French metrology institute [31,32] will eventually result in ultimate sub-Hz stabilities, accuracies and linewidths. One advantage of our method lies on the generation of a beat-note signal directly in the mid-IR without the need to pass the QCL beam through a sophisticated non-linear optical setup. As a result, much less than 1% of the QCL optical power is used for locking, allowing almost all of the QCL power to be exploited. We obtain a QCL that is both powerful and highly stable – potentially ultra-stable – to be exploited in a range of applications from fundamental research in quantum optics and high-resolution spectroscopy to quantum technologies or LiDAR. When driven at higher currents, our Fabry-Perot QCL operates in optical-frequency-comb regime (see supplementary materials). Our work thus also paves the way toward the development of powerful and fully stabilized metrology-grade QCL frequency combs.

**SUPPLEMENTARY MATERIAL**

See Supplementary material for supporting content.

**ACKNOWLEDGMENTS**

This work has been supported by the Region Ile-de-France in the framework of DIM SIRTEQ, by the Agence Nationale de la Recherche (project CORALI ANR-20-CE04-0006) and project COLECTOR ANR-19-CE30-0032) and by ENS-Thales Chair. The authors thank MenloSystems for fruitful discussions and for providing typical phase noise measurements of the OFCS repetition rate.

**REFERENCES**

[1] P. Cancio Pastor, I. Galli, G. Giusfredi, D. Mazzotti, and P. De Natale, "Testing the validity of Bose-Einstein statistics in molecules," Phys. Rev. A **92**(6), 063820 (2015).

[2] G. Barontini, L. Blackburn, V. Boyer, F. Butuc-Mayer, X. Calmet, J.R. Crespo López-Urrutia, E.A. Curtis, B. Darquié, J. Dunningham, N.J. Fitch, E.M. Forgan, K. Georgiou, P. Gill, R.M. Godun, J. Goldwin, V. Guarrera, A.C. Harwood, I.R. Hill, R.J. Hendricks, M. Jeong, M.Y.H. Johnson, M. Keller, L.P. Kozhiparambil Sajith, F. Kuipers, H.S. Margolis, C. Mayo, P. Newman, A.O. Parsons, L. Prokhorov, B.I. Robertson, J. Rodewald, M.S. Safronova, B.E. Sauer, M. Schioppo, N. Sherrill, Y.V. Stadnik, K. Szymaniec, M.R. Tarbutt, R.C. Thompson, A. Tofful, J. Tunesi, A. Vecchio, Y. Wang, and S. Worm, "Measuring the stability of fundamental constants with a network of clocks," EPJ Quantum Technol. **9**(1), 1–52 (2022).

[3] M.R. Fiechter, P.A.B. Haase, N. Saleh, P. Soulard, B. Tremblay, R.W.A. Havenith, R.G.E. Timmermans, P. Schwerdtfeger, J. Crassous, B. Darquié, L.F. Pašteka, and A. Borschevsky, "Toward Detection of the Molecular Parity Violation in Chiral Ru(acac)3 and Os(acac)3," J. Phys. Chem. Lett. **13**(42), 10011–10017 (2022).

[4] J. Lukusa Mudiayi, I. Maurin, T. Mashimo, J.C. de Aquino Carvalho, D. Bloch, S.K. Tokunaga, B. Darquié, and A. Laliotis, "Linear Probing of Molecules at Micrometric Distances from a Surface with Sub-Doppler Frequency Resolution," Phys. Rev. Lett. **127**(4), 043201 (2021).

[5] M.L. Diouf, F.M.J. Cozijn, B. Darquié, E.J. Salumbides, and W. Ubachs, "Lamb-dips and Lamb-peaks in the saturation spectrum of HD," Opt. Lett. **44**(19), 4733–4736 (2019).

[6] D.M. Segal, V. Lorent, R. Dubessy, and B. Darquié, "Studying fundamental physics using quantum enabled technologies with trapped molecular ions," Journal of Modern Optics **65**(5–6), 490–500 (2018).

[7] F. Benabid, F. Couny, J.C. Knight, T.A. Birks, and P.S.J. Russell, "Compact, stable and efficient all-fibre gas cells using hollow-core photonic crystal fibres," Nature **434**(7032), 488–491 (2005).

[8] D. Gatti, A.A. Mills, M.D. De Vizia, C. Mohr, I. Hartl, M. Marangoni, M. Fermann, and L. Gianfrani, "Frequency-comb-calibrated Doppler broadening thermometry," Phys. Rev. A **88**(1), 012514 (2013).

[9] J. Fischer, B. Fellmuth, C. Gaiser, T. Zandt, L. Pitre, F. Sparasci, M.D. Plimmer, M. de Podesta, R. Underwood, G. Sutton, G. Machin, R.M. Gavioso, D.M. Ripa, P.P.M. Steur, J. Qu, X.J. Feng, J. Zhang, M.R. Moldover, S.P. Benz, D.R. White, L. Gianfrani, A. Castrillo, L. Moretti, B. Darquié, E. Moufarej, C. Daussy, S. Briaudeau, O. Kozlova, L. Risegari, J.J. Segovia, M.C. Martín, and D. del Campo, "The Boltzmann project," Metrologia **55**(2), R1 (2018).

[10] E. Roueff, S. Sahal-Bréchot, M.S. Dimitrijević, N. Moreau, and H. Abgrall, "The Spectroscopic Atomic and Molecular Databases at the Paris Observatory," Atoms **8**(3), 36 (2020).

[11] I. Galli, S. Bartalini, R. Ballerini, M. Barucci, P. Cancio, M.D. Pas, G. Giusfredi, D. Mazzotti, N. Akikusa, and P.D. Natale, "Spectroscopic detection of radiocarbon dioxide at parts-per-quadrillion sensitivity," Optica **3**(4), 385–388 (2016).

[12] L. Tombez, S. Schilt, J. Di Francesco, T. Führer, B. Rein, T. Walther, G. Di Domenico, D. Hofstetter, and P. Thomann, in *Applied Physics B: Lasers and Optics* (2012), pp. 407–414.

[13] T.L. Myers, R.M. Williams, M.S. Taubman, C. Gmachl, F. Capasso, D.L. Sivco, J.N. Baillargeon, and A.Y. Cho, "Free-running frequency stability of mid-infrared quantum cascade lasers," Opt Lett **27**(3), 170–172 (2002).

[14] S. Bartalini, S. Borri, P. Cancio, A. Castrillo, I. Galli, G. Giusfredi, D. Mazzotti, L. Gianfrani, and P. De Natale, "Observing the Intrinsic Linewidth of a Quantum-Cascade Laser: Beyond the Schawlow-Townes Limit," Physical Review Letters **104**(8), 083904 (2010).

[15] L. Tombez, J. Di Francesco, S. Schilt, G. Di Domenico, J. Faist, P. Thomann, and D. Hofstetter, "Frequency noise of free-running 4.6 μm distributed feedback quantum cascade lasers near room temperature," Opt Lett **36**(16), 3109–3111 (2011).

[16] S. Bartalini, S. Borri, I. Galli, G. Giusfredi, D. Mazzotti, T. Edamura, N. Akikusa, M. Yamanishi, and P. De Natale, "Measuring frequency noise and intrinsic linewidth of a room-temperature DFB quantum cascade laser," Opt Express **19**(19), 17996–18003 (2011).

[17] F. Cappelli, G. Villares, S. Riedi, and J. Faist, "Intrinsic linewidth of quantum cascade laser frequency combs," Optica **2**, (2015).

[18] A.A. Mills, D. Gatti, J. Jiang, C. Mohr, W. Mefford, L. Gianfrani, M. Fermann, I. Hartl, and M. Marangoni, "Coherent phase lock of a 9 μm quantum cascade laser to a 2 μm thulium optical frequency comb," Opt Lett **37**(19), 4083–4085 (2012).

[19] I. Galli, M. Siciliani de Cumis, F. Cappelli, S. Bartalini, D. Mazzotti, S. Borri, A. Montori, N. Akikusa, M. Yamanishi, G. Giusfredi, P. Cancio, and P. De Natale, "Comb-assisted subkilohertz linewidth quantum cascade laser for high-precision mid-infrared spectroscopy," Appl. Phys. Lett. **102**(12), 121117 (2013).

[20] L. Consolino, F. Cappelli, M.S. de Cumis, and P.D. Natale, "QCL-based frequency metrology from the mid-infrared to the THz range: a review," Nanophotonics **8**(2), 181–204 (2019).

[21] F. Bielsa, A. Douillet, T. Valenzuela, J.-P. Karr, and L. Hilico, "Narrow-line phase-locked quantum cascade laser in the 9.2 μm range," Opt. Lett. **32**(12), 1641–1643 (2007).

[22] P.L.T. Sow, S. Mejri, S.K. Tokunaga, O. Lopez, A. Goncharov, B. Argence, C. Chardonnet, A. Amy-Klein, C. Daussy, and B. Darquié, "A widely tunable 10-μm quantum cascade laser phase-locked to a state-of-the-art mid-infrared reference for precision molecular spectroscopy," Applied Physics Letters **104**(26), 264101 (2014).

[23] F. Cappelli, I. Galli, S. Borri, G. Giusfredi, P. Cancio, D. Mazzotti, A. Montori, N. Akikusa, M. Yamanishi, and S. Bartalini, "Subkilohertz linewidth room-temperature mid-infrared quantum cascade laser using a molecular sub-Doppler reference," Optics Letters **37**(23), 4811–4813 (2012).

[24] S. Borri, I. Galli, F. Cappelli, A. Bismuto, S. Bartalini, P. Cancio, G. Giusfredi, D. Mazzotti, J. Faist, and P. De Natale, "Direct link of a mid-infrared QCL to a frequency comb by optical injection," Optics Letters **37**(6), 1011–1013 (2012).

[25] A. Gambetta, M. Cassinerio, N. Coluccelli, E. Fasci, A. Castrillo, L. Gianfrani, D. Gatti, M. Marangoni, P. Laporta, and G. Galzerano, "Direct phase-locking of a 8.6-μm quantum cascade laser to a mid-IR optical frequency comb: application to precision spectroscopy of $N_2O$," Opt. Lett. **40**(3), 304–307 (2015).

[26] S. Borri, S. Bartalini, T. Leveque, L. Gianfrani, P.D. Natale, P. Cancio, G. Giusfredi, D. Mazzotti, and I. Galli, "Frequency-comb-referenced quantum-cascade laser at 4.4 μm," Optics Letters **32**(8), 988–990 (2007).

[27] A. Gambetta, E. Vicentini, Y. Wang, N. Coluccelli, E. Fasci, L. Gianfrani, A. Castrillo, V.D. Sarno, L. Santamaria, P. Maddaloni, P.D. Natale, P. Laporta, and G. Galzerano, "Absolute frequency measurements of $CHF_3$ Doppler-free ro-vibrational transitions at 8.6 μm," Opt. Lett. **42**(10), 1911–1914 (2017).

[28] A. Gambetta, E. Fasci, A. Castrillo, M. Marangoni, G. Galzerano, G. Casa, P. Laporta, and L. Gianfrani, "Frequency metrology in the near-infrared spectrum of $H_2^{17}O$ and $H_2^{18}O$ molecules: testing a new inversion method for retrieval of energy levels," New Journal of Physics **12**(10), 103006 (2010).

[29] M.G. Hansen, E. Magoulakis, Q.-F. Chen, I. Ernsting, and S. Schiller, "Quantum cascade laser-based mid-IR frequency metrology system with ultra-narrow linewidth and $1 \times 10^{-13}$-level frequency instability," Opt. Lett., OL **40**(10), 2289–2292 (2015).

[30] G. Insero, S. Borri, D. Calonico, P.C. Pastor, C. Clivati, D. D'Ambrosio, P. De Natale, M. Inguscio, F. Levi, and G. Santambrogio, "Measuring molecular frequencies in the 1–10 μm range at 11-digits accuracy," Sci Rep **7**(1), 12780 (2017).

[31] R. Santagata, D.B.A. Tran, B. Argence, O. Lopez, S.K. Tokunaga, F. Wiotte, H. Mouhamad, A. Goncharov, M. Abgrall, Y. Le Coq, H. Alvarez-Martinez, R. Le Targat, W.K. Lee, D. Xu, P.-E.P.-E. Pottie, B. Darquié, and A. Amy-Klein, "High-precision methanol spectroscopy with a widely tunable SI-traceable frequency-comb-based mid-infrared QCL," Optica **6**(4), 411 (2019).

[32] B. Argence, B. Chanteau, O. Lopez, D. Nicolodi, M. Abgrall, C. Chardonnet, C. Daussy, B. Darquié, Y. Le Coq, and A. Amy-Klein, "Quantum cascade laser frequency stabilization at the sub-Hz level," Nature Photonics **9**(7), 456–460 (2015).

[33] D.S. Elliott, R. Roy, and S.J. Smith, "Extracavity laser band-shape and bandwidth modification," Phys. Rev. A **26**(1), 12–18 (1982).

[34] M. Zhu, and J.L. Hall, "Stabilization of optical phase/frequency of a laser system: application to a commercial dye laser with an external stabilizer," JOSA B **10**(5), 802–816 (1993).

# Highly coherent phase-lock of an 8.1 μm quantum cascade laser to a turn-key mid-IR frequency comb: supplemental document.

1- Comb-like operation of the QCL laser

The QCL used in this work is a Fabry-Perot source grown and processed at ETH Zürich. Its active region is made of a strain compensated dual stack heterostructure grown by molecular bean epitaxy. The 3 mm long device has a high reflectivity (HR) coating ($Al_2O_3$/Au) on its back facet and is electrically injected with a low-noise current driver (Model QubeCL from ppqSense). This QCL operates at a temperature of 263 K at which the threshold current is 400 mA for a maximum emitted power in continuous wave of 350 mW (see Fig. S1(a)). For the locking demonstration, the laser is operated in a single mode oscillating at a wavelength of ~8.160 μm. It is operated close to threshold for a current ranging from 400 to 460 mA and delivers up to 45 mW. Upon further increase of the driving current, many modes start oscillating and the device self-stabilizes a frequency comb-like characterized by a narrow intermodal beating at a free spectral range frequency of 14.8 GHz starting from a current of 540 mA (see Fig.S1(b)). The intermodal beat-note shown in Fig.S2 is measured via a bias-T that sends the RF part of the laser current to a spectrum analyzer.

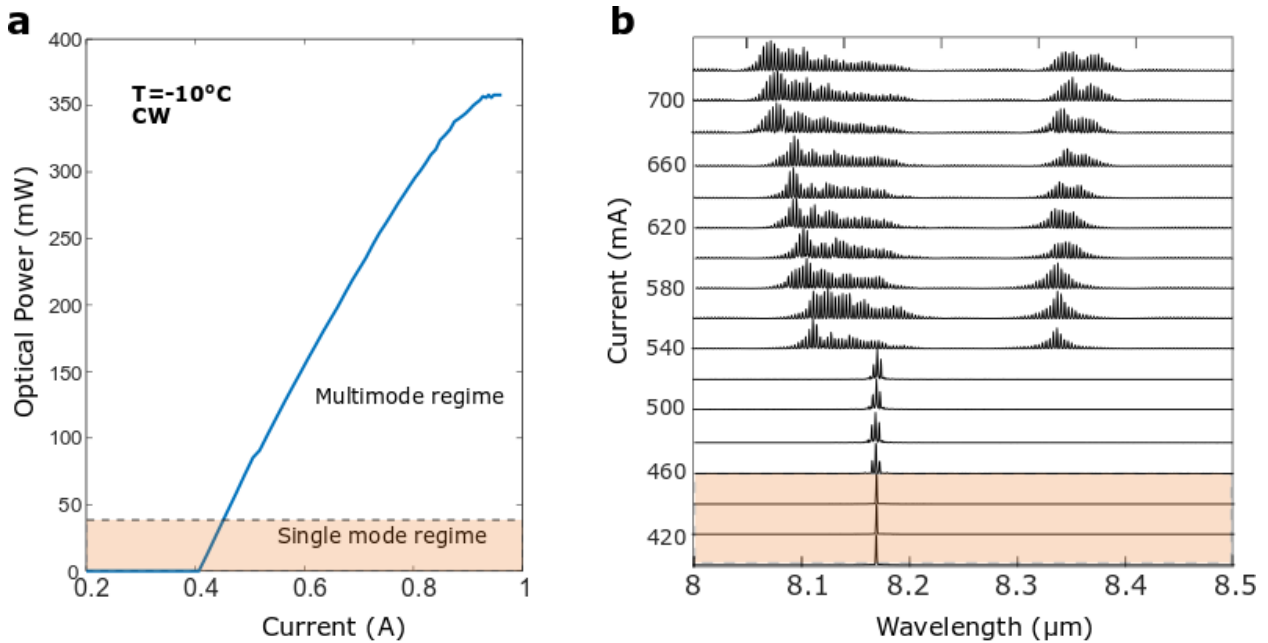

Fig. S1 (a) Power-current curve of the QCL (3 mm long, standard HR coating on the back facet) in operation at 263 K. Single mode and multimode regimes are highlighted. (b) Optical spectra obtained at 263 K for different driving currents measured with an optical spectrum analyzer (Bristol instruments, 2GHz of resolution).

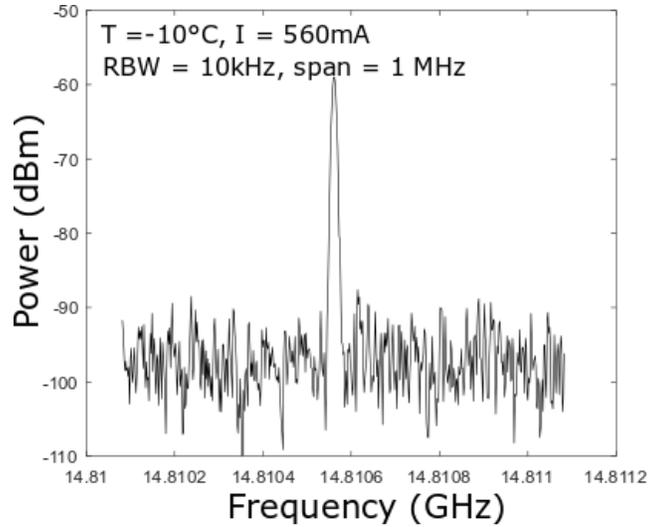

Fig. S2 Intermodal beat-note of the laser operating at T=-10°C and I = 560 mA (RBW = 10 kHz and span = 1 MHz).

2- Heterodyne signal and detector saturation

In this section we derive the signal to noise ratio of our detection setup. The noise density (in V²/Hz) at the output of our mercury-cadmium-telluride (MCT) detector (VIGO System PVI-4TE-10.6) is dominated by the thermal noise of its equivalent resistance given by:

$$S_N = 4\frac{k_b T}{R_d}R_f^2 = 1.9 \times 10^{-14}\ V^2/Hz$$

Where $R_d$ is close to 50 Ω and $R_f = 8.65 \times 10^3$ V/A is the gain of the amplifier (FIP-1k-1G-F-M4-ND) and T=200 K.

The heterodyne signal is generated from the beating between the signal mode QC laser with a power of $P_{qcl} = 50\mu W$ and one tooth of the mid-IR optical frequency comb synthesizer (OFCS) with an overall OFCS mean power of 500µW resulting in $P_{OFCS} = 10\ nW$ per tooth. The amplitude of the heterodyne signel (in V) at the output of the amplifier becomes $I_{het} = 2R\sqrt{P_{qcl}P_{OFCS}} \times R_f = 6.1\ mV$ with R = 0.5 A/W is the effective responsivity of the detector. This gives us a $SNR_{1Hz} = 10\log\left(\frac{I_{het}^2/2}{S_N \times 1Hz}\right) = +90dB$ in good agreement with the experimental value.

The saturation of the detector amplifier (FIP-1k-1G-F-M4-ND, 800MHz bandwidth) occurs when the signal amplitude at the output is >0.5 V. In order to avoid this staturation due the strong oscillations of the Mid-IR OFCS at the harmonics of $f_{rep}$ = 100MHz , we performed an optical filtering using a 50 GHz resolution monochromator. We thus reduce the number of optical comb modes by a factor of ~84. After optical filtering, the maximum amplitude of the signal oscillating at $f_{rep}$ at the output of the detector is given by :

$$I_{OFCS} = N_{rf} \times [RR_f 2N_{opt}P_{OFCS}] = \sim 0.346\ V$$

where $N_{opt}$ is the number modes in the optical bandwidth (about 43 000 in the full bandwidth but only about 500 after spectral filtering) and $N_{rf}$ the number of oscillations in the amplifier bandwidth (about 800MHz/100MHz = 8 here).